# 3D CA model of tumor-induced angiogenesis

Monjoy Saha#, Amit Kumar Ray*, Swapan Kumar Basu$
#,* School of Biomedical Engineering, IIT- Banaras Hindu University
$ Department of Computer Science, Banaras Hindu University
# monjoybme@gmail.com
* amit.ray@rediffmail.com
$ swapankb@gmail.com

**Abstract —***Tumor-induced angiogenesis is the formation of new sprouts from preexisting nearby parent blood vessels. Computationally, tumor-induced angiogenesis can be modeled using cellular automata (CA), partial differential equations, etc. In this present study, a realistic physiological approach has been made to model the process of angiogenesis by using 3D CA model. CA technique uses various neighborhoods like Von-Neumann neighborhood, Moore neighborhood, and Margolus neighborhood. In our model Von-Neumann neighborhood has used for distribution of some significant chemical and non-chemical tumor angiogenic factors like vascular endothelial growth factor, endothelial cells, $O_2$, extracellular matrix, fibronectin, etc., and Moore neighborhood is used for distribution of matrix metalloproteinase. In vivo tumor environment all the factors are not distributed equally in the extracellular matrix. Distributions of those chemical and nonchemical factors depend on their source, nature and function. To keep similarity with the biological tumor environment, we have formulated initial distributions of the chemical and non-chemical factors accordingly. We have started the simulation in MATLAB with this initial distribution. Number of sprouts randomly varies from one run to another. We observed that sprouts are not originating from the same locations in each simulation. A sprout has high sensitivity of VEGF and fibronectin concentrations. sVEGFR-1 always tries to regress the sprout. When two or more sprouts come closer, they merge with each other leading to anastomosis. Sufficient number of tip cells may cause sprout towards tumor.*

## I. INTRODUCTION

Tumor is the uncontrolled growth of cells. At the initial stage of tumor, nutrients and oxygen are supplied by the surrounding extracellular matrix through passive diffusion [1]. Requirement of nutrients and oxygen increase when the tumor diameter exceeds 2mm [2]. In that situation nutrients and oxygen can't reach at the center of the tumor. Cells of that region completely die, that region is called **necrotic core**. The cells of the outer region of the necrotic core may remain alive if those cells get sufficient nutrients and oxygen. This region is called the quiescent zone. The outer region of the tumor is called the proliferating zone. The cells of this region get nutrients and oxygen from the surrounding extracellular matrix. Figure 1 shows the three-layer structure of the tumor.

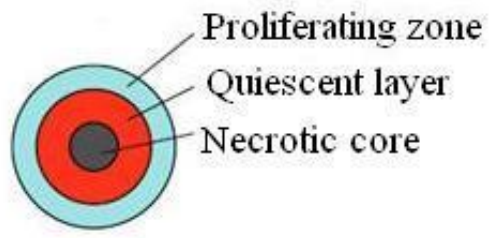

Fig.1: *Three-layer structure of a tumor spheroid*

Oxygen and nutrients can only diffuse 0.1 to 0.2 mm of tumor. Tumor growth is divided into three phases: first phase is the avascular phase, second phase is angiogenesis, and the third phase is the vascular phase [3]. Deficiency of oxygen in the tumor causes tumor hypoxia. In tumor hypoxia, hypoxia inducible factor (HIF) upregulates secretion of some tumor angiogenic factors (TAFs) such as Fibroblast growth factor (FGF), angiopoietins (Ang1 and Ang2), vascular endothelial growth factor (VEGF), transforming growth factor (TGF), platelet derived growth factor (PDGF), etc. VEGF, the main important tumor angiogenic factor [4][5][6], is a glycoprotein with molecular weight 43–46 kDa. VEGF also acts as an endothelial cell mitogen. VEGF activation causes tip and stalk cell formation. These cells proliferate and migrate along the opening of the basement membrane for new sprout formation. Matrix metalloproteinases (MMPs) expressed by the tip cells clear the extracellular matrix (ECM). MMPs expression decreases when stalk cells come into contact with pericytes, which are assigned by the PDGF. Transforming growth factor (TGF) activates when endothelial cells interact with mural cells. TGF helps in basement membrane production in the newly formed blood vessel. Figure 2 shows the steps of tumor-induced angiogenesis.

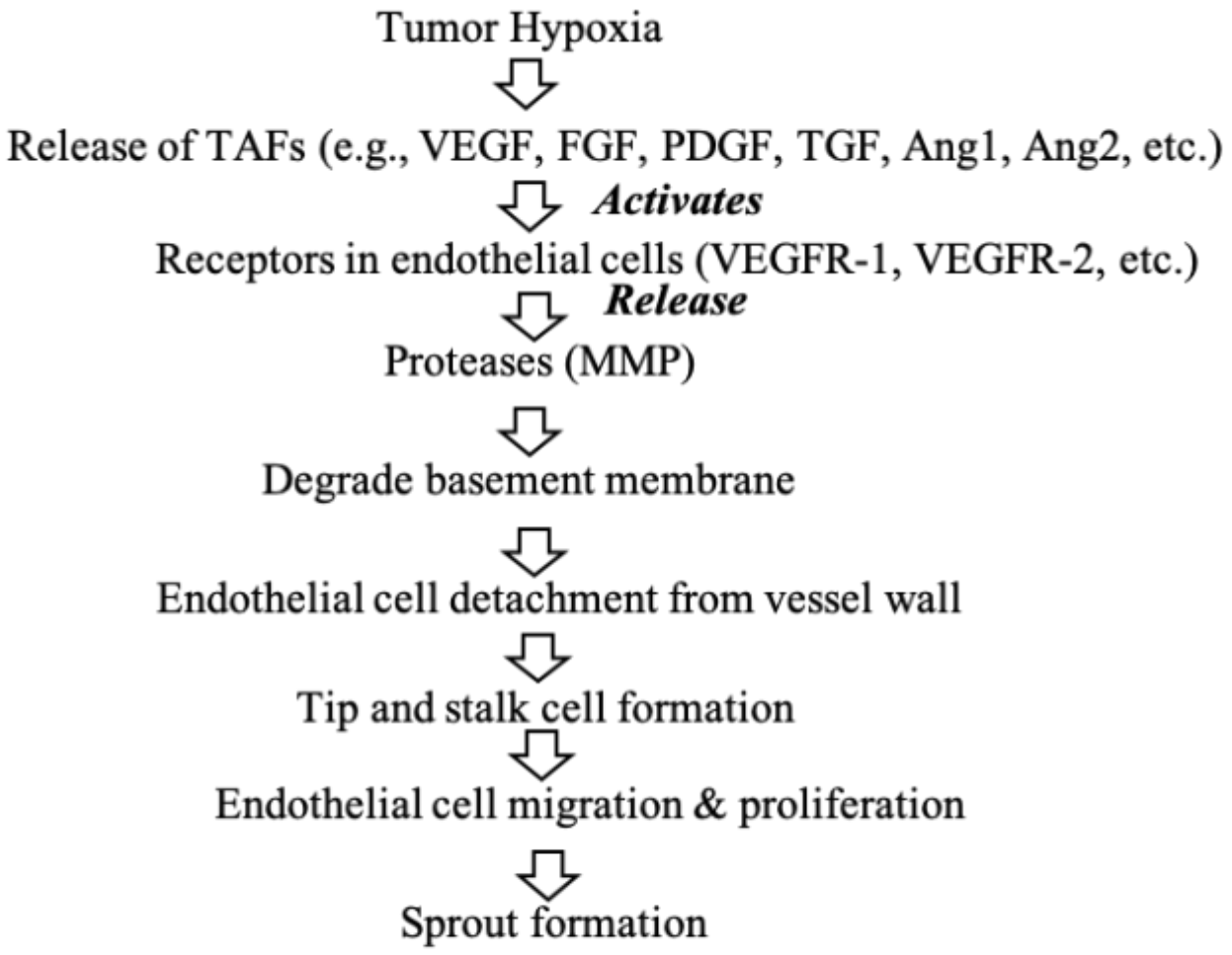

Fig.2. Steps of tumor-induced angiogenesis

This paper has been organized as an introduction, formulation and simulation, results and discussion, and conclusion.

## II. FORMULATION AND SIMULATION

This model uses concentration of VEGF, fibronectin (F), soluble VEGF receptor-1 also known as sVEGFR-1, endothelial cells (ECs), MMP, oxygen ($O_2$), and ECM. The concentrations of all the factors mentioned above are normalized from 0 to 1. The tissue is represented by a grid or cells of cellular automata. We have used two types of neighborhoods of 3D cellular automata. Figure 3(a) shows 2D Von-Neumann neighborhood and Figure 3(b) shows 2D Moore neighborhood.

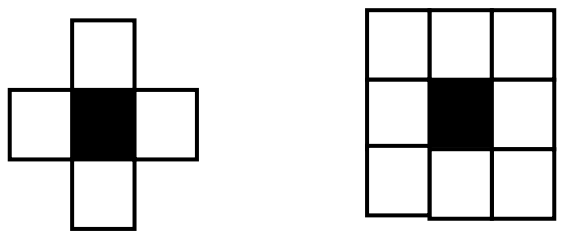

(a) (b)

Fig.3 (a) 2D Von Neumann, and (b) 2D Moore neighborhoods

The concentrations of an entity in a cell at time ($t$ + 1) depend on the concentrations of the relevant quantities of its neighbors at time $t$.

The twenty six neighbors of the cell at [i] [j] [k] in 3D Cartesian domain are

[i+1][j][k],[i-1][j][k],[i][j+1][k],[i+1][j+1][k],[i-1][j+1][k], [i-1][j-1][k],[i][j-1][k],[i+1][j-1][k],[i][j][k+1], [i+1][j][k+1],[i-1][j][k+1],[i][j+1][k+1],[i+1][j+1][k+1],[i-1][j+1][k+1],[i-1][j-1][k+1],[i][j-1][k+1],[i+1][j-1][k+1], [i][j][k-1],[i+1][j][k-1],[i-1][j][k-1],[i][j+1][k-1], [i+1][j+1][k-1],[i-1][j+1][k-1],[i-1][j-1][k-1],[i][j-1][k-1], [i+1][j-1][k-1].

Time evolution for all the factors (except MMP) is given by Equation (1).

$$S_{(i,j,k)(t+1)} = f(S_{(i+1,j,k)(t)} + S_{(i-1,j,k)(t)} + S_{(i,j+1,k)(t)} + S_{(i+1,j+1,k)(t)} + S_{(i-1,j+1,k)(t)} + S_{(i-1,j-1,k)(t)} + S_{(i,j-1,k)(t)} + S_{(i+1,j-1,k)(t)} + S_{(i,j,k+1)(t)} + S_{(i+1,j,k+1)(t)} + S_{(i-1,j,k+1)(t)} + S_{(i,j+1,k+1)(t)} + S_{(i+1,j+1,k+1)(t)} + S_{(i-1,j+1,k+1)(t)} + S_{(i-1,j-1,k+1)(t)} + S_{(i,j-1,k+1)(t)} + S_{(i+1,j-1,k+1)(t)} + S_{(i,j,k-1)(t)} + S_{(i+1,j,k-1)(t)} + S_{(i-1,j,k-1)(t)} + S_{(i,j+1,k-1)(t)} + S_{(i+1,j+1,k-1)(t)} + S_{(i-1,j+1,k-1)(t)} + S_{(i-1,j-1,k-1)(t)} + S_{(i,j-1,k-1)(t)} + S_{(i+1,j-1,k-1)(t)}) \quad (1)$$

Here, $S_{(i,j,k)(t+1)}$ is the state of the cell ($i, j, k$) at time $t$+1. All the grid elements are updated concurrently. The simulation uses cellular automata to model 3D tumor angiogenesis; at each time step, the entire tissue is updated by simulating VEGF concentration, endothelial cell density, sVEGFR-1 concentration, fibronectin concentration, MMP,
concentration in the presence of oxygen and nutrients. Time evolution for MMP is given by Equation (2).

$$M^{t+1}_{(i,j,k)} = M^{t}_{(i+1,j,k)} + M^{t}_{(i-1,j,k)} + M^{t}_{(i,j+1,k)} + M^{t}_{(i,j-1,k)} + M^{t}_{(i,j,k+1)} + M^{t}_{(i+1,j,k+1)} + M^{t}_{(i-1,j,k+1)} + M^{t}_{(i,j+1,k+1)} + M^{t}_{(i,j-1,k+1)} + M^{t}_{(i,j,k-1)} + M^{t}_{(i+1,j,k-1)} + M^{t}_{(i-1,j,k-1)} + M^{t}_{(i,j+1,k-1)} + M^{t}_{(i,j-1,k-1)} \quad (2)$$

**Local Rules for Tumor Angiogenesis**

An endothelial cell migrates to a new site, if sufficient number of endothelial cells is present along with adequate quantities of VEGF, fibronectin, MMP, oxygen. sVEGFR-1 may (in very less amount) or may not be present. An EC can proliferate and migrate to a new site, if VEGF and fibronectin concentrations exceed threshold values. For capillary branching, the following conditions are applicable:

i. Parent sprout must be older than the age of daughter sprout.
ii. For the generation of new sprouts, there should be enough space.
iii. To generate new sprouts from the existing sprout tips, sufficient number of ECs near the sprout tip needs to be present.
iv. The density of ECs is inversely proportional to the concentration of VEGF.
v. The possibility of generating new sprouts increases with the concentration of VEGF.

The direction of tip cells is dependent on VEGF concentration, fibronectin concentration, MMP concentration, ECM concentration, oxygen, endothelial cell density.

**Distribution of model variables**

*A. Distribution of VEGF in the extracellular matrix at various time steps:*

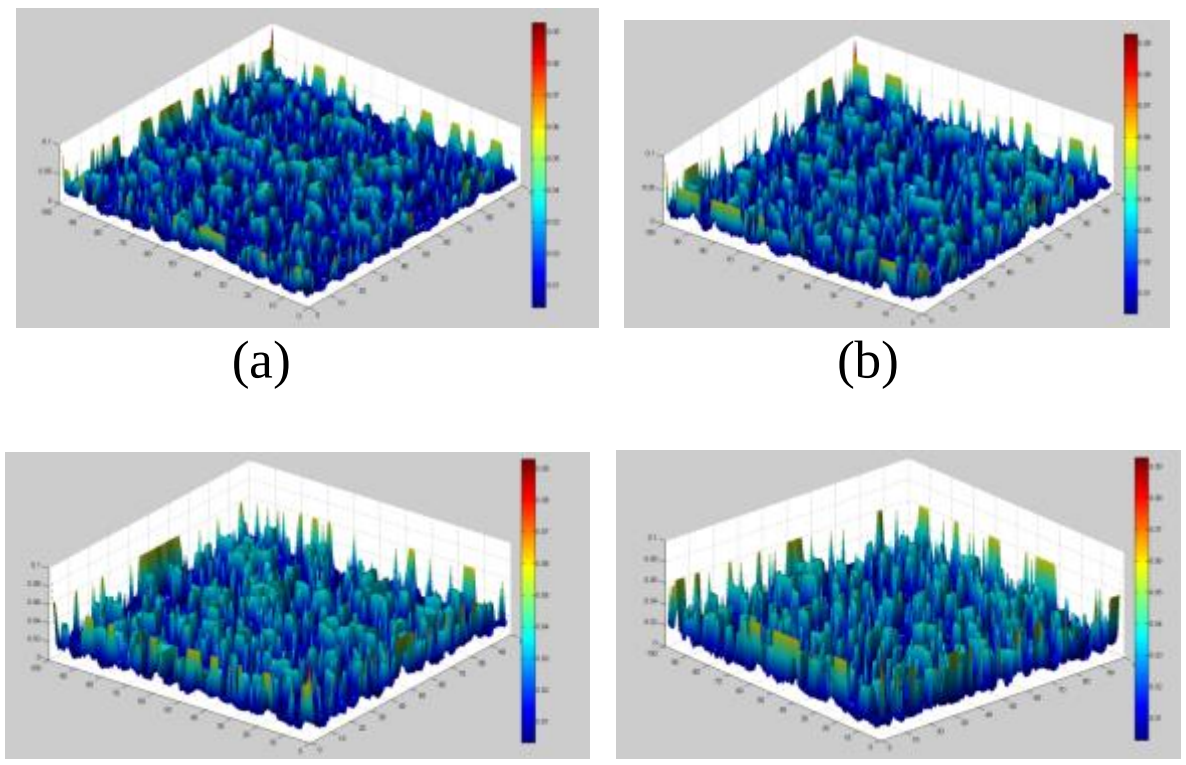

(a) (b) (c) (d)

Fig. 4: *VEGF concentration in extracellular matrix after* 25, 50, 75, 100 *steps simulation respectively.*

Figure 4 shows VEGF concentration at different time steps. In each time of simulation, it is found that VEGF concentration is high at the tumor site and gradually diffuses towards blood vessels. In each simulation concentration level varies.

*B. Distribution of fibronectin in the extracellular matrix*

Figure 5 shows Fibronectin distribution in the extracellular matrix at various time steps. Fibronectin are secreted from endothelial cells and some fibronectin are degraded by extracellular matrix, so its distribution is somewhat different.

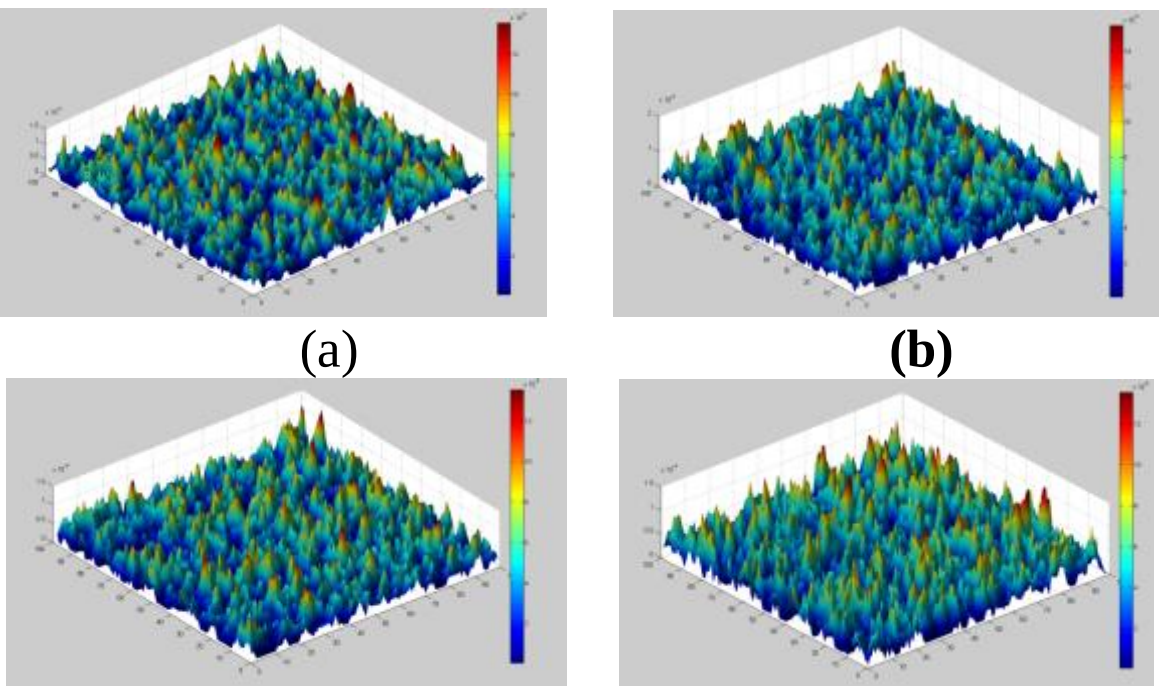

(a) **(b)**

**(c)** **(d)**

Fig. 5: *Fibronectin concentration in extracellular matrix after* 25, 50, 75, 100 *steps of simulation respectively.*

*C. Distribution of sVEGFR-1 in the extracellular matrix*

Figure 6 shows sVEGFR-1 concentration after 25, 50, 75, 100 steps simulation. It is evident that sVEGFR-1 concentration gradually decreases and VEGF concentration increases. Due to this reason newly formed blood vessels extend towards tumor.

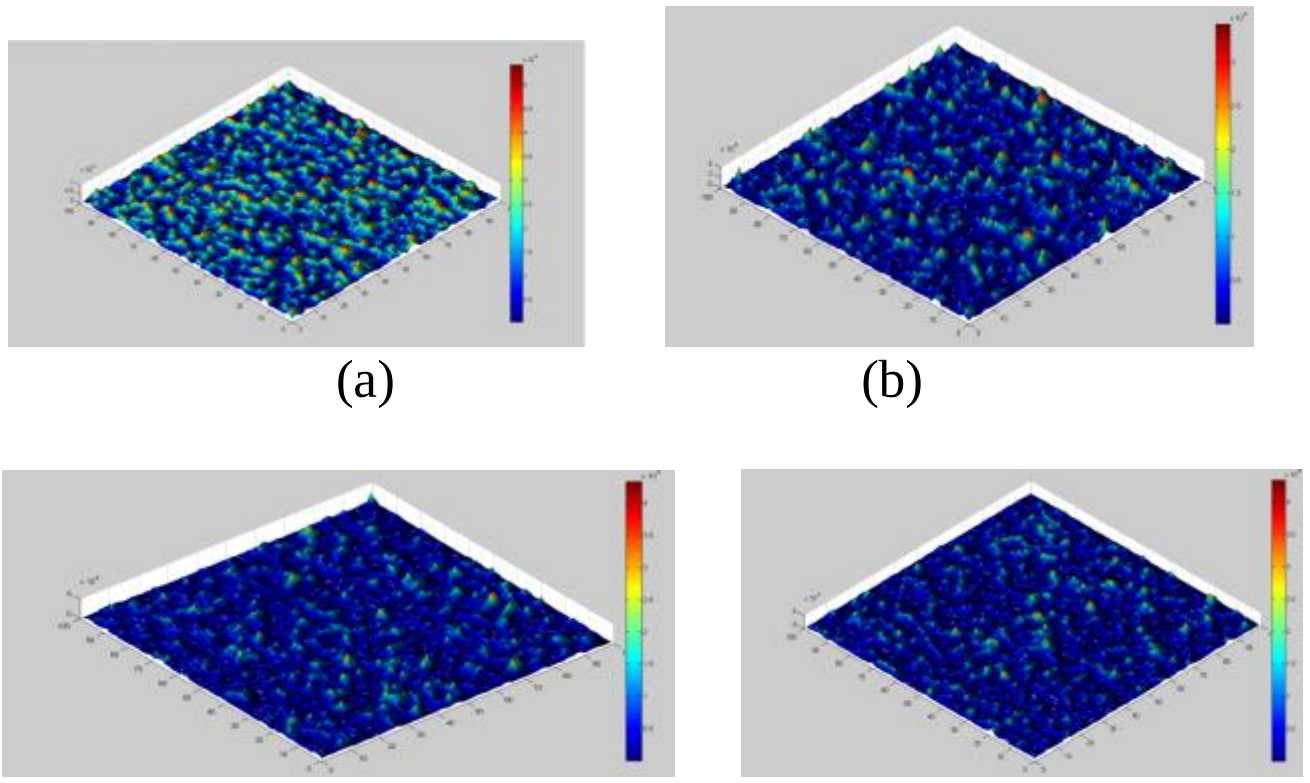

(a) (b)

(c) (d)

Fig. 6: *sVEGFR-1 concentration after* 25, 50, 75, 100 *steps simulation respectively.*

*D. Distribution of MMP in the extracellular matrix*

Figure 7 shows matrix metalloprotease concentration after 25, 50, 75, 100 steps of simulation respectively. Distribution of matrix metalloprotease is based on Moore neighborhood.

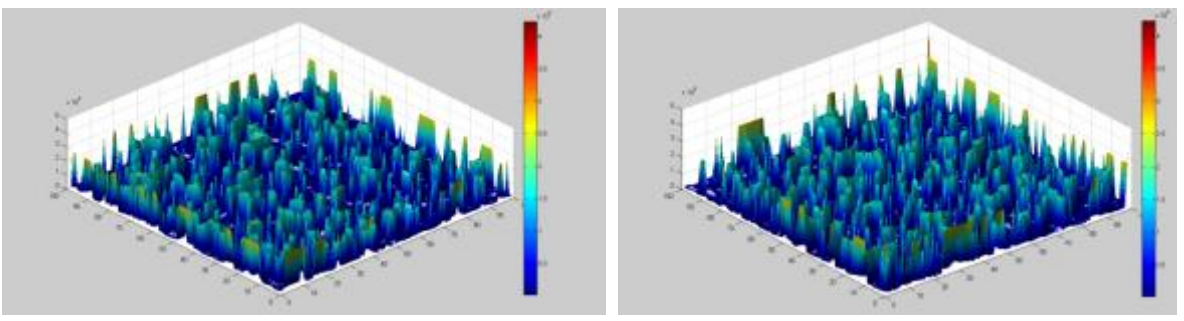

(a) (b)

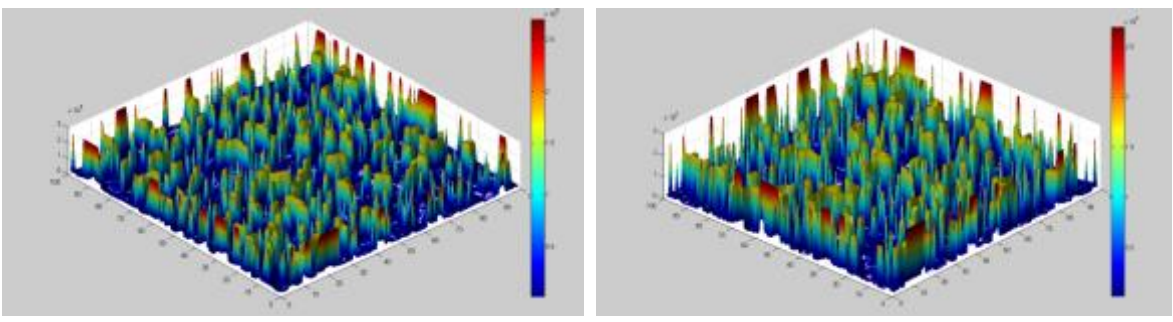

(c) (d)

Fig. 7: *Matrix metalloproteinase concentration after* 25, 50, 75, 100 *steps simulation respectively.*

*E. Distribution of extracellular matrix*

Figure 8 shows extracellular matrix concentration at different points of the grid.

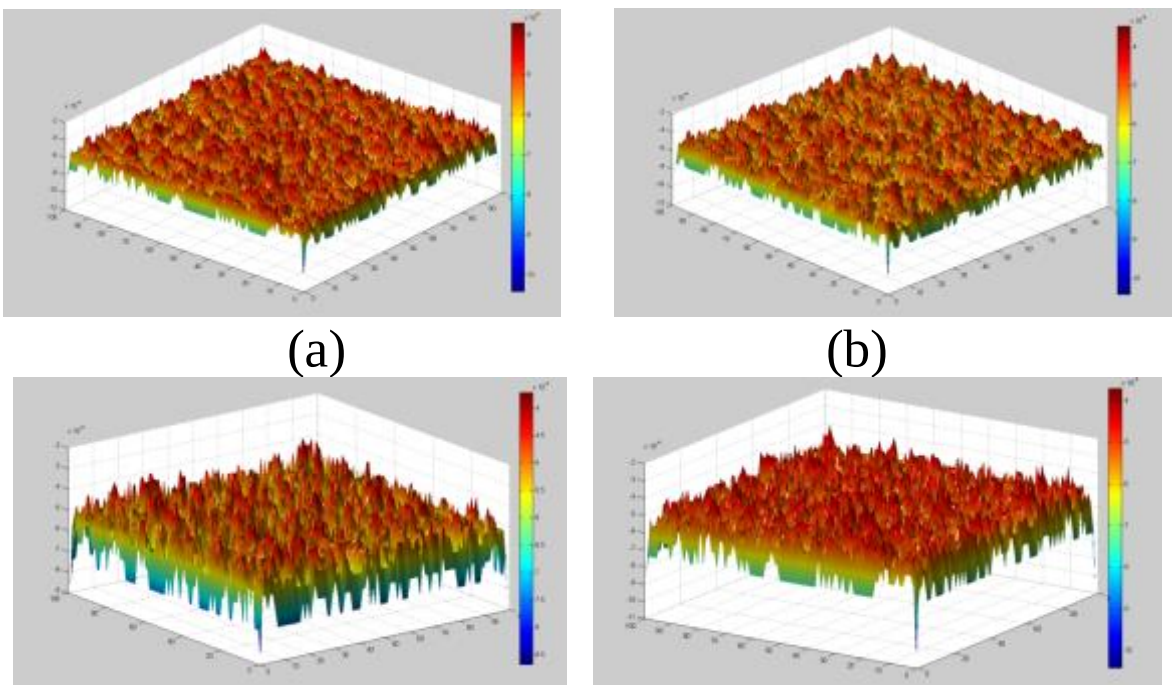

(a) (b)

**(c)** **(d)**

Fig.8: *Extracellular matrix concentration after* 25, 50, 75, 100 *steps of simulation respectively.*

*F. Distribution of nutrients & oxygen in the extracellular matrix*

Figure 9 shows Nutrients and oxygen concentration.

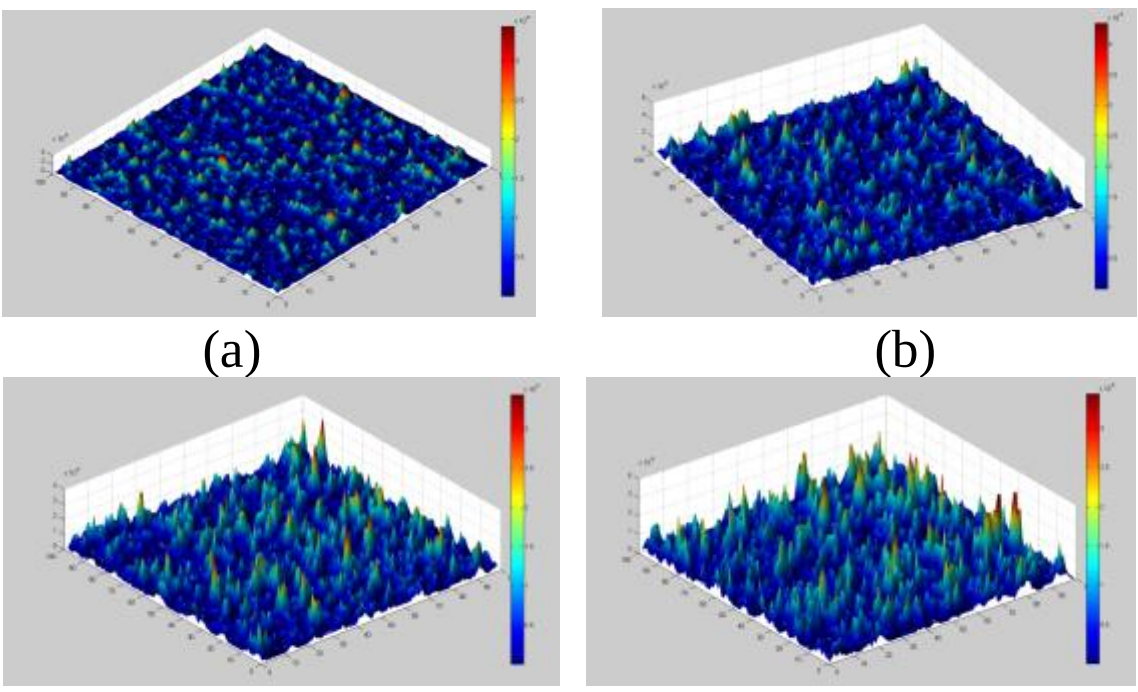

(a) (b)

(c) (d)

Fig.9: *Nutrients & Oxygen concentration in extracellular matrix after* 25, 50, 75, 100 *steps of simulation respectively.*

## III. RESULTS AND DISCUSSION

Our 3D cellular automata model is able to predict a variety of sprout formation patterns. In this model VEGF and fibronectin concentration plays an important role in sprout formation. sVEGFR-1 concentration helps in restricting the sprout formation. We obtained a number of sprout patterns by variation of different parameters. This model has been developed by keeping in mind some significant biological phenomena of the tumor environment such as endothelial cell proliferation and interactions with ECM, anastomosis, vessel maturation, sprout branching, secretion of angiogenic factors from a solid tumor.

The model was implemented in MATLAB 7.4.0 (R2007a) as a sequential program and run on personal laptop in Windows 7 operating system with 2GB RAM.

### *A. Blood vessels extending at various time steps*

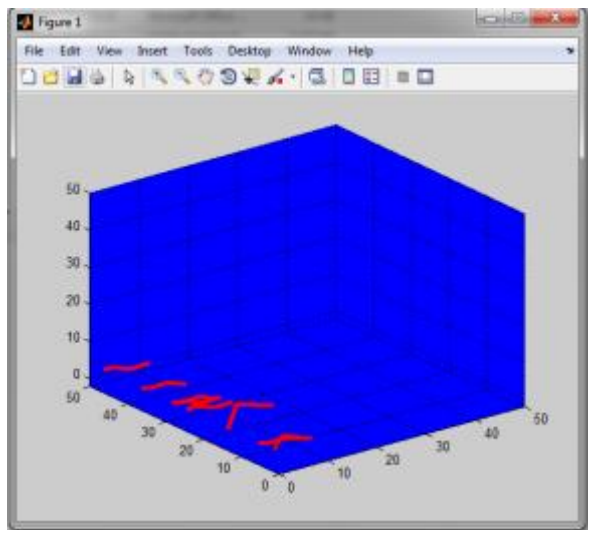

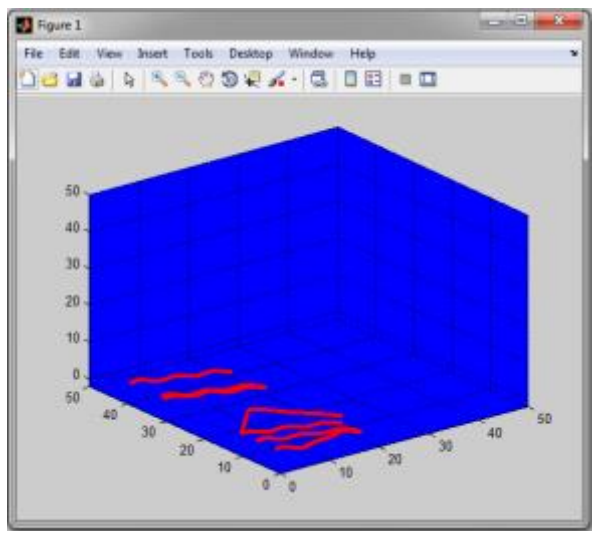

T=25 T=50

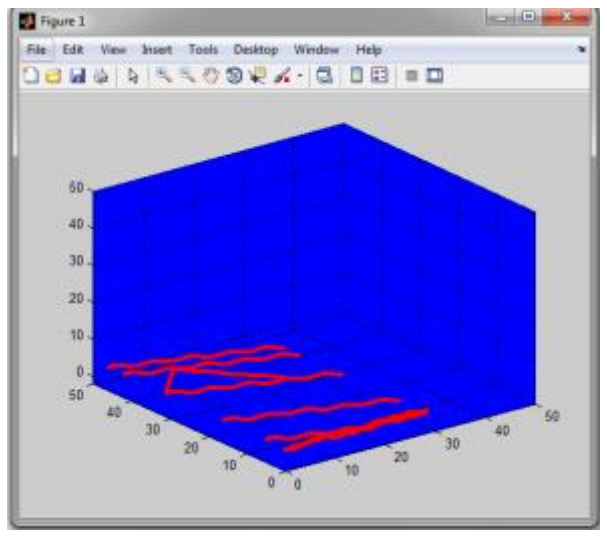

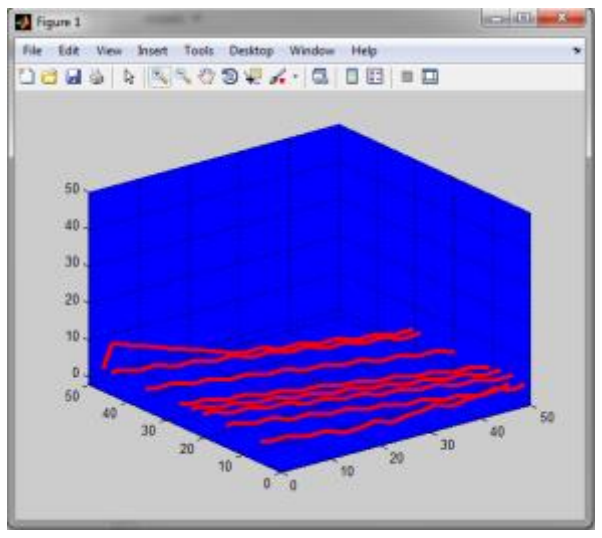

T=75 T=100

Fig.9 (a): *Blood vessel extension at various time steps*.

Figure 9 shows extension of new sprouts at various time steps. The direction of the tip cell follows a higher VEGF concentration gradient. Sprouts are originating from the parent vessels, situated at the left side along *j* axis and tumor is in the opposite wall of *j* axis, from different locations and at different angles. After 25 steps of simulation, sprouts are originating from different locations, but they are not going in the same direction. Some sprouts are going to the extreme left direction, some going towards extreme right. We have increased the simulation time to 50, 75, and 100. In each case we found that cell density at the tumor site increases in each simulation. As the cell density increases the number of sprouts also increases. When two or more sprouts come closer, they fuse to form anastomosis.

### *B. Effects of VEGF on tumor-induced angiogenesis*

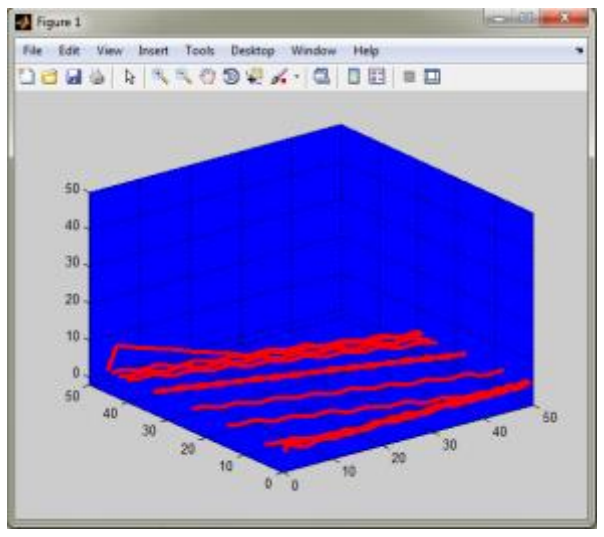

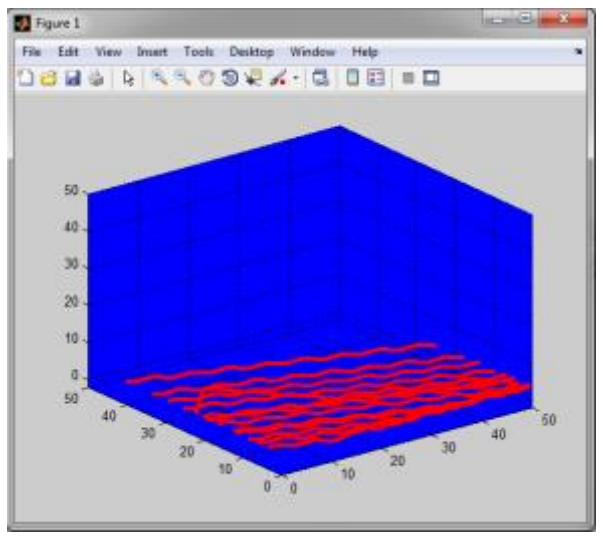

(a) (b)

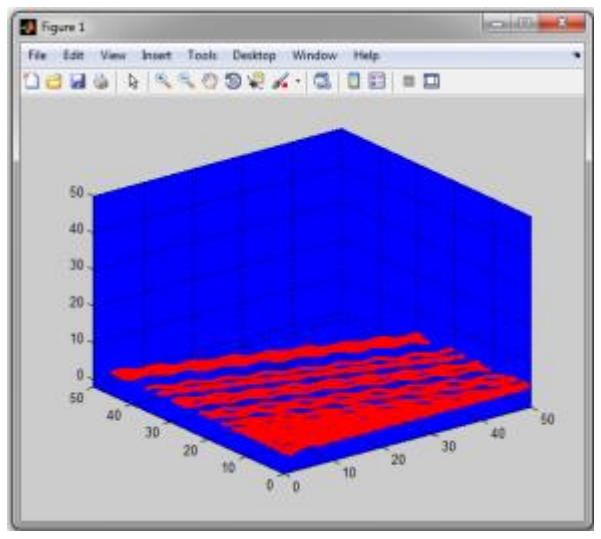

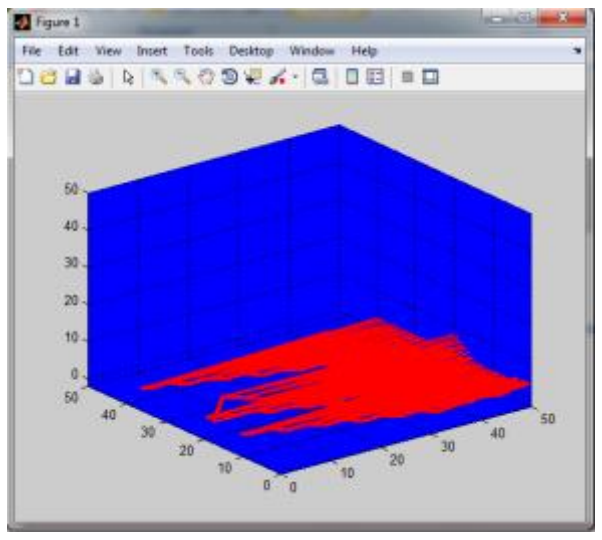

(c) (d)

Fig.10: *Effects of VEGF in tumor-induced angiogenesis.*

Figure 10 shows effects of VEGF in tumor-induced angiogenesis. Here we simulate the program by changing VEGF concentration from normal to highest value. At each simulation the number of sprouts increases. This also causes more anastomosis. In Figure 5 (d), there is highest VEGF concentration. In this situation it is also found that more sprouts may originate from the existing sprouts.

### *C. Effects of sVEGFR-1in sprout formation*

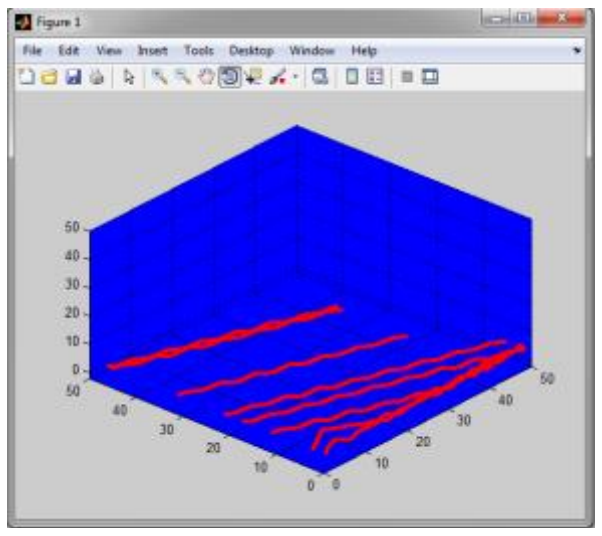

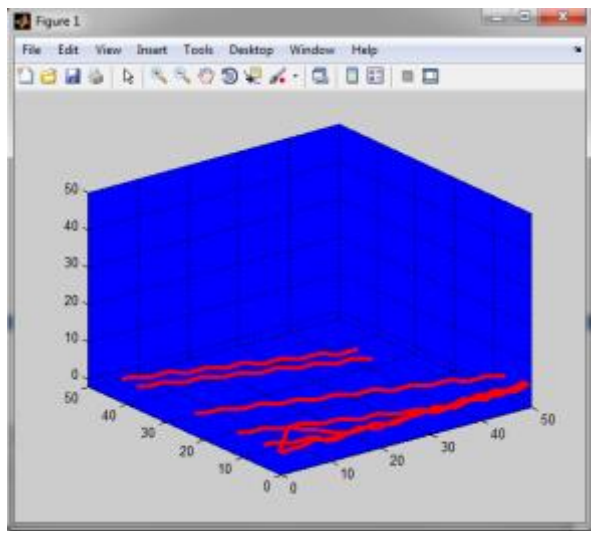

(a) (b)

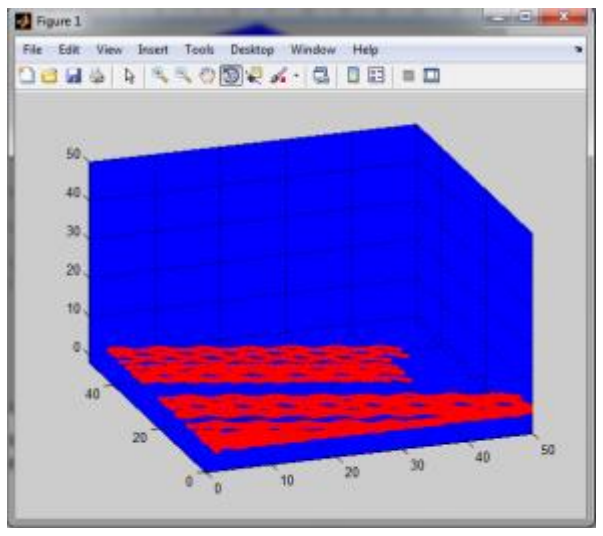

(c)

Fig. 11**:** *Effects of sVEGFR-1 in tumor-induced angiogenesis*

Figure 11 shows Effects of sVEGFR-1 in tumor-induced angiogenesis. We increased sVEGFR-1 concentration gradually and found that sVEGFR-1 restricts the formation of new sprouts. With the increased presence of sVEGFR-1, some sprouts can’t reach the tumor. The main reason behind this is that tip cells of those blood vessels lose proliferation capacity, because of very low VEGF concentration. In this case tip cells may die later. sVEGFR-1 acts as the main antitumor agent

### *D. Effects of fibronectin on sprout formation*

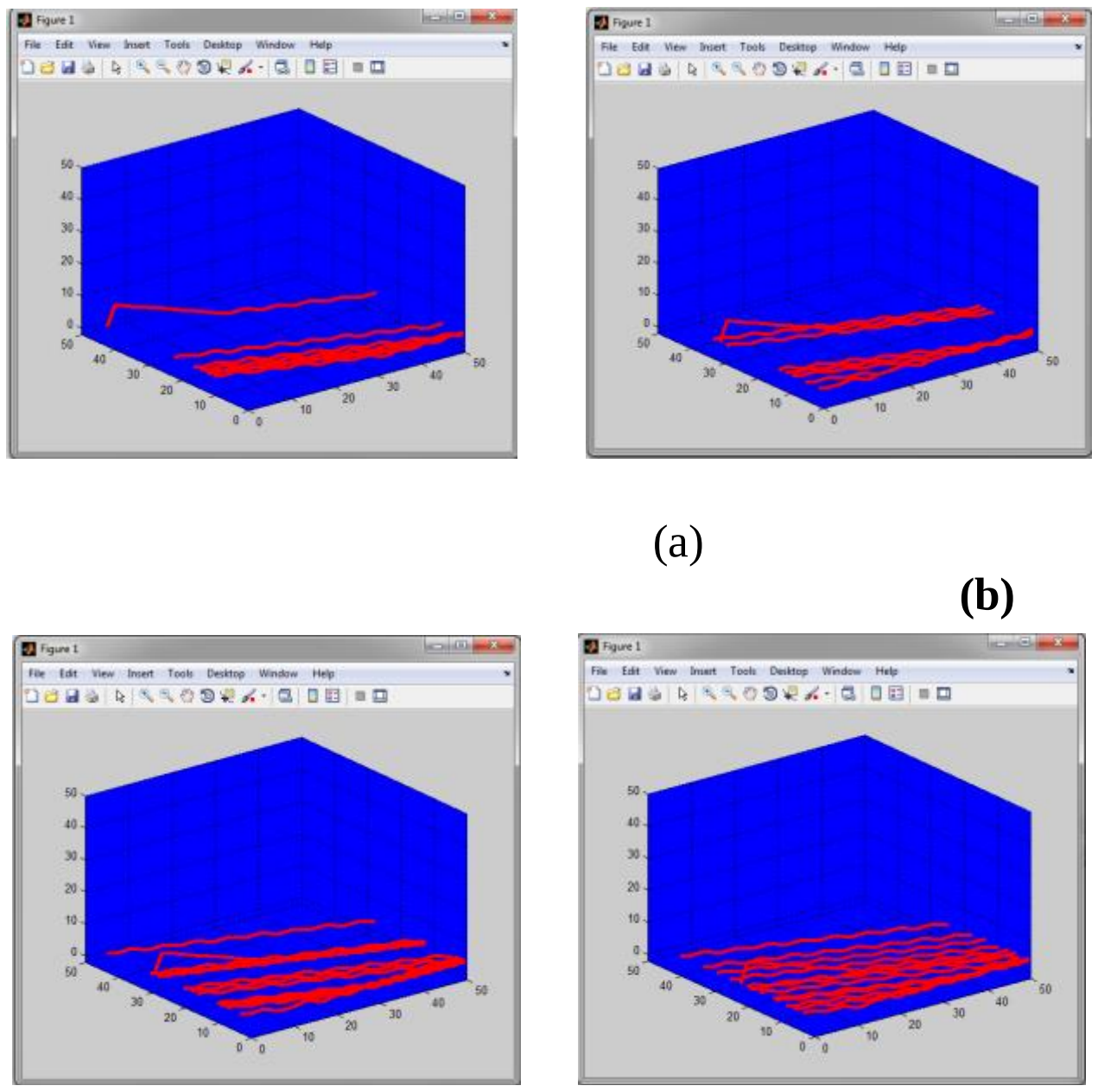

(a) **(b)**

(c) (d)

Fig.12: *Effects of fibronectin tumor-induced angiogenesis*

Figure 12 shows effects of fibronectin in tumor-induced angiogenesis. We gradually increased fibronectin concentration from normal value to highest value. In each case we have found that the number of sprouts and anastomosis increases. If there is no fibronectin present in the extracellular matrix, no sprout is formed. Fibronectin acts as a catalyst for VEGF.

*E. Effects of cell density in tumor-induced angiogenesis*

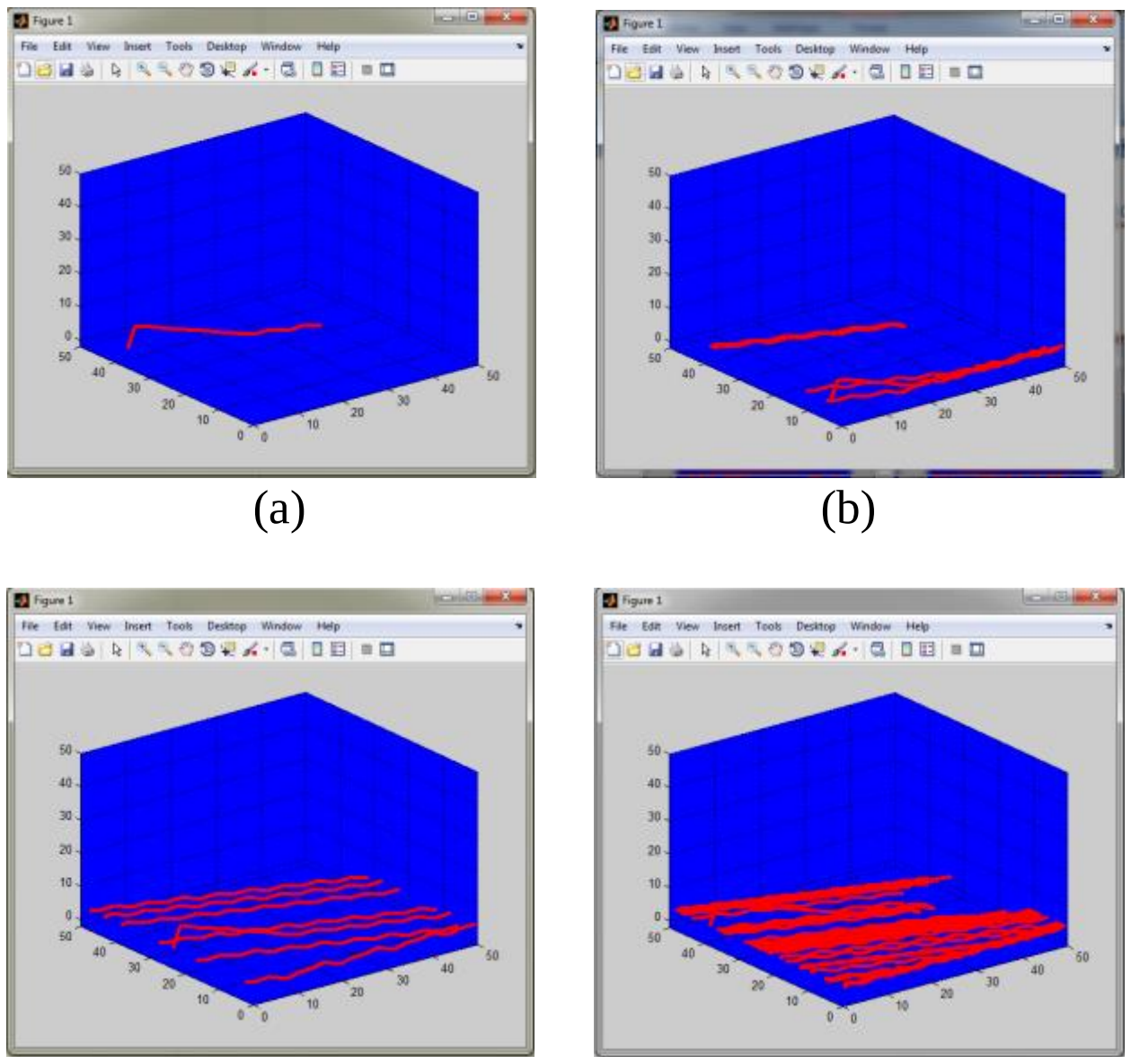

(a) (b)

(c) (d)

Fig.13: *Effects of cell density in tumor-induced angiogenesis*

Figure 12 shows effects of cell density in tumor-induced angiogenesis. Here we are gradually increasing endothelial cell density in the parent vessel. Initially endothelial cell density in the parent vessel is very low, so we are getting one sprout originating from the parent vessel. In the next simulation when endothelial cell density is more, the number of sprouts is also more.

## IV. CONCLUSION

The final results of 3D Cellular Automata is promising. Using local interactions in 3D tissue space, study of the effects of various biological factors like VEGF, (sVEGFR-1, matrix metalloproteinase (MMP), fibronectin, etc. on tumor vasculature has been carried out. Some of the parameters in this model vary randomly. In each run of simulation concentrations of the different chemical and non-chemical entities varies. The number of sprout formation, extension of sprouts and occurrence of anastomosis also vary from run to run. We did not find a huge difference between VEGF and fibronectin action. Endothelial cell density increases towards the tumor. sVEGFR-1 resists the action of VEGF. So, we can use sVEGFR-1 as an anti-tumor agent. We also found that a sufficient number of endothelial cells may create tip and sprout. Otherwise sprout formation is not possible.

The simulation results, though in the preliminary stage, are promising. Further study needs to be carried out involving more variables representing other biological actors. One can use other modeling techniques for the problem for cross verification. The authors would like to continue further investigation in this direction.

***Acknowledgement: Few changes have been made from the original manuscript, which was published in ICMSDPA, 2012 proceeding (offline).***